\documentclass[conference]{IEEEtran}
\IEEEoverridecommandlockouts
\usepackage{cite}
\usepackage{amsmath,amssymb,amsfonts}
\usepackage{algorithmic}
\usepackage{graphicx}
\usepackage{textcomp}
\usepackage{xcolor}
\usepackage{caption}

\def\BibTeX{{\rm B\kern-.05em{\sc i\kern-.025em b}\kern-.08em
    T\kern-.1667em\lower.7ex\hbox{E}\kern-.125emX}}
\begin{document}
\graphicspath{{./Figures/}}
	\title{Improving SAGIN Resilience to Jamming with Reconfigurable Intelligent Surfaces}

    \author{ \IEEEauthorblockN{ Leila Marandi\IEEEauthorrefmark{1}, Khaled Humadi\IEEEauthorrefmark{1}, Gunes Karabulut-Kurt\IEEEauthorrefmark{1},  Wessam Ajib\IEEEauthorrefmark{2}, Wei-Ping Zhu\IEEEauthorrefmark{3}}
    
    \IEEEauthorblockA{\IEEEauthorrefmark{1} Department of Electrical Engineering, Polytechnique Montreal, Montreal, Canada},
    \IEEEauthorblockA{\IEEEauthorrefmark{2} Department of Computer Sciences, University of Quebec in Montreal, Montreal, Canada},
    \IEEEauthorblockA{\IEEEauthorrefmark{3} Department of Electrical and Computer Engineering, Concordia University, Montreal, Canada.}
    \IEEEauthorblockA{
        Email: \{leila.marandi, khaled.humadi, gunes.kurt\}@polymtl.ca, ajib.wessam@uqam.ca, 
        weiping@ece.concordia.ca.
        % weiping.zhu@concordia.ca.
    }
    }

% \author{\IEEEauthorblockN{Leila Marandi}   
% \IEEEauthorblockA{\textit{dept. of electrical engineering} \\
% \textit{Polytechnique Montreal}\\
% Montreal, Canada \\
% leila.marandi@polymtl.ca}
% \and
% \IEEEauthorblockN{Khaled Humadi}
% \IEEEauthorblockA{\textit{dept. of electrical engineering} \\
% \textit{Polytechnique Montreal}\\
% Montreal, Canada \\
% khaled.humadi@polymtl.ca}
% \and
% \IEEEauthorblockN{Gunes Karabulut Kurt}
% \IEEEauthorblockA{\textit{dept. of electrical engineering} \\
% \textit{Polytechnique Montreal}\\
% Montreal, Canada \\
% gunes.kurt@polymtl.ca}
% \and 
% \IEEEauthorblockN{Wessam Ajib}
% \IEEEauthorblockA{\textit{dept. of Computer Sciences} \\
% \textit{University of Quebec in Montreal}\\
% Montreal, Canada \\
%  ajib.wessam@uqam.ca}
% \and 
% \IEEEauthorblockN{Wei-Ping Zhu}
% \IEEEauthorblockA{\textit{dept.Electrical and Computer Engineering} \\
% \textit{Concordia University}\\
% Montreal, Canada \\
% weiping@ece.concordia.ca}
% }
\maketitle
\bstctlcite{IEEEexample:BSTcontrol} % for references:: show author names instead dash

\begin{abstract} 

This study investigates the anti-jamming space-air-ground integrated network (SAGIN) scenario wherein a reconfigurable intelligent surface (RIS) is deployed on a fixed Unmanned Aerial Vehicle (UAV) to counteract malevolent jamming attacks. In contrast to existing research, in this paper, we consider that a Low Earth Orbit (LEO) satellite is sending the signal to the user on the ground in the presence of jamming from a Geostationary Equatorial Orbit (GEO) satellite side. We aim to maximize the signal-to-jamming plus noise ratio (SJNR) by optimizing the RIS beamforming and transmit power of the LEO satellite. Assuming the availability of global channel state information (CSI) at the RIS, we propose alternating optimization (AO) and semidefinite relaxation (SDR) techniques to address the complexity. Simulation results show that the optimization schemes lead to considerable performance improvements.
% Additionally, it has been confirmed that the RIS enhances link quality compared to systems lacking an RIS. 
The results also indicate that, given the high jamming power and the relatively small number of RIS elements, deploying the RIS on UAVs near the user is more effective in mitigating the impact of jamming interferers. 
% than placing it on a high-altitude platform (HAP).

\end{abstract}

% Geostationary Equatorial Orbit (GEO), Low Earth Orbit (LEO), Unmanned Aerial Vehicle (UAV)

\section{Introduction} \label{sec:intro}
\begin{figure}[t] 
    \centering
    \includegraphics[width=0.8\linewidth, height=9cm, keepaspectratio=false]{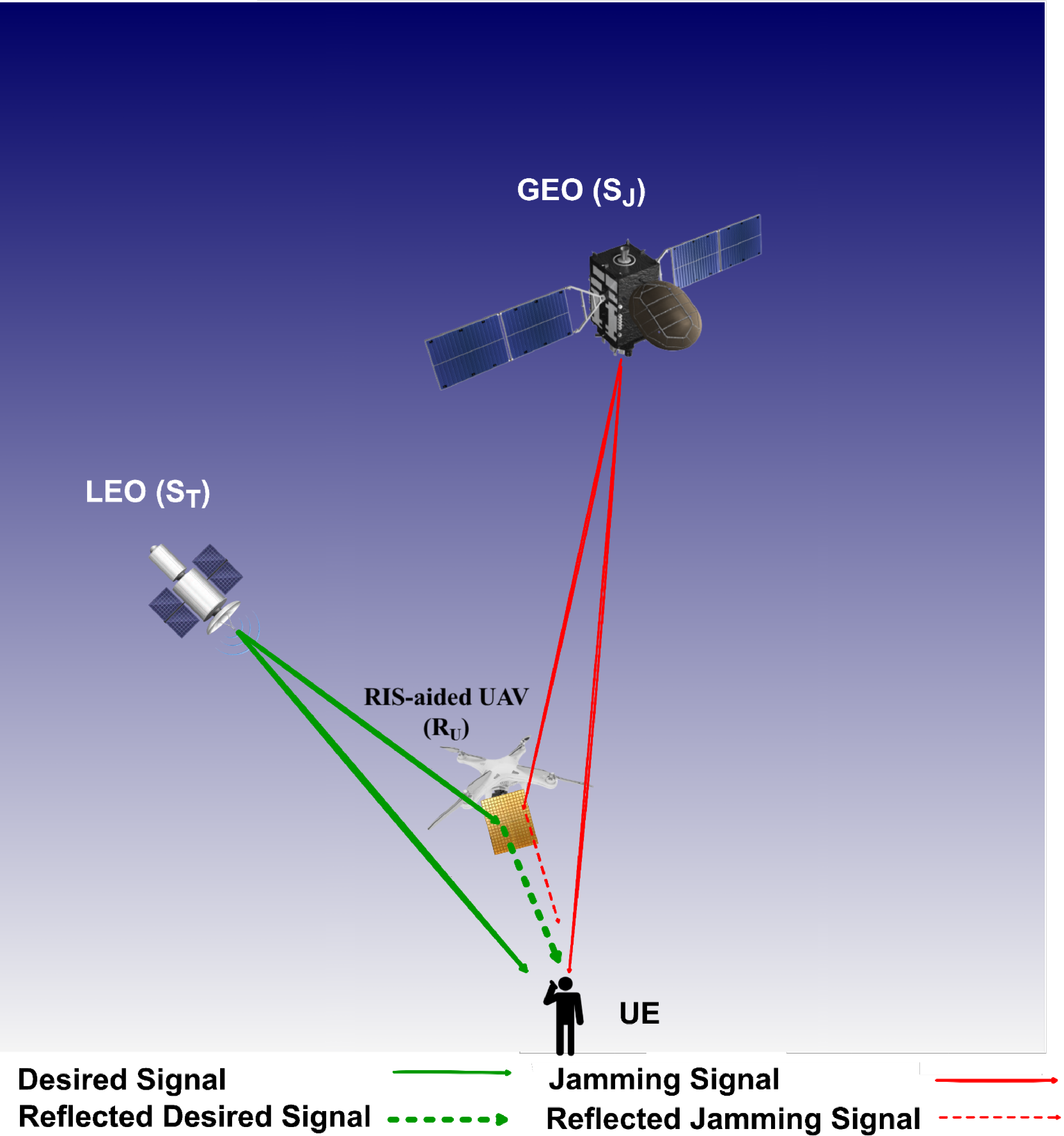} %Cropped_GEO_LEO_UAV_v2.eps
    \caption{\scriptsize System model consisting of GEO as a jammer ($S_J$), LEO as a transmitter ($S_T$), and RIS mounted on a fixed UAV ($R_U$).}
    \label{fig: system_model}
\end{figure}

Given the rapid development of applications, the increase in the number of users, and the demand for high data rates, next-generation wireless networks need to support high spectral efficiency and extensive connectivity \cite{Energy_walid}. Providing effective support through terrestrial networks in scenarios such as maritime surveillance, counter-communication actions, etc faces challenges. Moreover, over 80\% of the world including lands and oceans are not covered by terrestrial networks. 
In this context, satellites play a vital role in the networks of the sixth generation (6G), playing a key role in reducing traffic congestion and improving the efficiency of resource allocation while expanding their communication coverage  \cite{Anti_Jamming_NOMA}. 
Although line-of-sight (LoS) satellite communication can increase capacity and quality of service across wide areas, shadowing, and obstructions make it challenging to sustain LoS connection between the satellite and terrestrial users \cite{RIS_UAV_Interference}.\\
The emergence of Space - Air - Ground Integrated Networks (SAGIN) has significantly transformed the landscape of global communication systems. SAGIN, offering a comprehensive network architecture, integrates space, air, and terrestrial communication layers to ensure seamless and ubiquitous connectivity across various domains. This integration, supporting a wide range of applications from remote sensing and environmental monitoring to urban planning and emergency response systems, highlights the importance and critical role of SAGIN in advancing communication capabilities \cite{SAGSIN_Kishk}. However, the transmission efficiency of SAGINs is especially vulnerable to significant free-space loss, shadowing phenomena, and the damaging effects of jamming attacks due to the broadcast nature of wireless transmission and the prevalent line-of-sight (LoS) connections. These factors collectively limit communication quality \cite{Anti_Jamming_NOMA, Joint_Trajectory_RIS_UAV}. \\
In addition to the well-known advantages of SAGIN, they can also be subject to large-scale attacks \cite{Jamming_Khaled}. The possibility of broad jamming is one of the main concerns regarding hacking of decommissioned satellites, as the incident with the Canadian Anik F1R satellite. These satellites, originally positioned in Geostationary Equatorial Orbit (GEO) to provide extensive coverage, can inadvertently become powerful tools for broadcasting unauthorized signals over a broad area when commandeered \cite{dead_satellite, Decommissioned_Satellite, Broadcast_Hacker}. This capability means that a single hacked satellite can affect ground users across a vast region, leading to significant disruptions in communication networks. To mitigate the risks associated with satellite jamming, several solutions are being explored and implemented. One common method is using frequency hopping spread spectrum (FHSS), which rapidly switches the transmission frequency to avoid jammers. This makes it harder for the jammer to lock onto the signal, but it may still suffer from synchronization issues and limited performance in highly dynamic environments \cite{FHSS_Zhang, FHSS_Zhao, jamming_attack_pirayesh}. Another widely used approach is direct-sequence spread spectrum (DSSS), which spreads the signal over a wider bandwidth to make it more resistant to interference. However, DSSS is vulnerable to wideband jamming and code-matching attacks if the jammer has knowledge of the spreading code \cite{jamming_attack_pirayesh}. Dynamic spectrum access (DSA) is a more advanced solution that uses AI algorithms to monitor spectrum usage and switch to cleaner frequencies. This technique can effectively avoid jamming, but it requires real-time spectrum sensing and fast decision-making, which adds complexity and processing delay \cite{DSA_wang}. The other method is beamforming and MIMO-based techniques that use multiple antennas to focus the signal in the direction of the legitimate receiver and direct nulls toward the jammer. While they can significantly reduce the impact of jamming, these methods depend on complex hardware and accurate channel estimation \cite{MIMO_Jin, jamming_attack_pirayesh, MIMO_Lang}. Moreover, error correction coding can help recover data in the presence of interference by adding redundancy to the signal. This can improve reliability but introduces extra delay and becomes ineffective if the signal is heavily jammed \cite{jamming_attack_pirayesh}. 
Overall, while these methods provide varying levels of protection, most of them require changes to hardware, signal processing, or spectrum management.
% One of the approaches is adding encryption devices to payloads. This method ensures that even if a satellite is compromised, the data it carries remains protected from unauthorized access or tampering \cite{Encryption_data_satellite}. Another effective approach is frequency hopping, which rapidly changes transmission frequencies to evade jamming signals, making it difficult for the jammer to lock onto a specific frequency \cite{Freq-hop, RL-Frequ-hop}. The other strategy is the use of spread spectrum technology, where the signal is spread over a wide frequency range. This method enhances the signal's resilience against jamming by making it harder for the jammer to affect the entire spectrum \cite{Magazine-AntiJam, Anti-Spectrum}. Additionally, directional antennas play a crucial role in mitigating jamming. These antennas focus the communication beam in specific directions, thereby reducing susceptibility to interference from jammers located outside the main beam path \cite{AntiJam-DirectionAntenna}. Furthermore, adaptive filtering is employed to further enhance communication security. Advanced filtering techniques are used to distinguish between legitimate signals and jamming interference, allowing the network to filter out the unwanted jamming signals effectively\cite{GNSS_Anti_Jamming}.\\
% % One of the approaches is adding encryption devices to payloads. This method ensures that even if a satellite is compromised, the data it carries remains protected from unauthorized access or tampering \cite{Encryption_data_satellite}.
A method that has attracted significant attention in recent years is employing Reconfigurable Intelligent Surface (RIS)  \cite{Joint_Trajectory_RIS_UAV, Anti-RIS_UAV}. Unlike other methods, which focus on obscuring or spreading signals, RIS significantly enhance communication systems by dynamically altering signal paths to optimize signal strength and reduce interference, including jamming \cite{MultiLayer_RIS_Game_Hierarchical, Jamming_mitigation_RIS_Passive_Beamforming}. Additionally, recent surveys emphasize the value of RIS in physical-layer security, emphasizing its function in safe transmission across 6G networks and jamming mitigation \cite{Jury3_ref1, Jury3_ref2}. Ground networks through optimizing RIS elements independently to improve anti-jamming performance and reduce computational complexity. Similarly, \cite{AntiJamming_RIS_DQN} addresses improving anti-jamming communication by employing solar-powered RIS in tactical wireless sensor networks, utilizing a deep Q network-based algorithm to optimize energy allocation and RIS phase shifts. \cite{AntiJamming_RIS_Game} and \cite{MultiLayer_RIS_Game_Hierarchical} both utilize game theory approaches to model interactions between base stations and jammers, with \cite{AntiJamming_RIS_Game} framing it as a Bayesian Stackelberg game and \cite{MultiLayer_RIS_Game_Hierarchical} employing a hierarchical game-learning approach to reach equilibrium under practical constraints like discrete RIS phases and imperfect channel state information. The authors in \cite{Jamming_mitigation_RIS_Passive_Beamforming} explore the potential of aerial RIS  for anti-jamming, focusing on optimal deployment and passive beamforming, while \cite{Joint_Space_Freq_antiJamming_RIS} introduce a joint space-frequency scheme to balance flexibility and effectiveness in combatting jamming signals, indicating enhanced resistance with an increased number of RIS elements suitable for military and civilian uses.

The above mentioned articles demonstrate the variety and efficiency of RIS in anti-jamming strategies, offering a flexible and powerful tool that can enhance network resilience.
\begin{table}[t]
	\captionsetup{font={scriptsize, color=black}}
	\begin{center}
		\caption{\MakeUppercase{SUMMARY OF NOTATION}}
		\label{table:Notation}
		\begin{tabular}{|l| |l|}
			\hline	
			\textbf{Notation} & \textbf{Description}\\ \hline \hline
			${S_T}, {S_J}, {R_U}$ &
			\begin{tabular}{l}
				 Notation of LEO satellite as a BS, \\ Jammer,  RIS-UAV assisted node 
			\end{tabular}\\ \hline	
   
			${{\mathbf{C}}_{{{\mathrm{S}}_T}}}, {{\mathbf{C}}_{{{\mathrm{S}}_J}}},{{\mathbf{C}}_{UE}}, {{\mathbf{C}}_{R_U}} $ &
			\begin{tabular}{l}
				3D Cartesian coordinates of the BS, \\Jammer, UE, and UAV \\  
			\end{tabular}  \\ \hline
            $K = {K_r}\times{K_c}$ &
            \begin{tabular}{l}
				Number of reflecting elements \\in RIS
			\end{tabular} \\ \hline
			${h_{_{{S_T}-UE}}}, {h_{_{{S_J}-UE}}}$& 
            \begin{tabular}{l}
				 Direct channel gain  between \\satellites (LEO, GEO) and UE\\  
			\end{tabular}\\ \hline
			${{h_{_{{S_{T}}-{R_U}-UE}}}}, {{h_{_{{S_{J}}-{R_U}-UE}}}}$ & 
            \begin{tabular}{l}
				 Cascade channel gain between \\ satellites  (LEO, GEO), RIS and UE\\  
			\end{tabular}\\ \hline            
			$\rho$ &
			\begin{tabular}{l} 
				The path loss at the reference \\distance $1$ $m$
			\end{tabular} \\ \hline
			${g_{_{{S_T}-UE}}}, \bold{g_{_{{S_T}-{R_U}}}}$ & \begin{tabular}{l} 
				Phase response of  the $\text S_T$-UE \\ and $\text S_T$-RIS channel
			\end{tabular} \\ \hline
			$d, d_{_{{S_T}-UE}}, d_{_{{S_T}-{R_U}}}$ &
			\begin{tabular}{l}
				 The separation of the RIS element,\\  Distance between $\text S_T$ and UE,\\ Distance between $\text S_T$ and RIS.
			\end{tabular} \\ \hline
			${\phi _{_{{S_T}-{R_U}}}}, {\varphi _{_{{S_T}-{R_U}}}}$ & 
			\begin{tabular}{l}
				The angle of arrival in vertical\\ and horizontal from ${S_T}$ \\
			\end{tabular}\\ \hline
			${{P_{{{{S_T}}}}}}, {{P_{{{{S_J}}}}}}$ &\begin{tabular}{l}
			 The transmit power of the BS\\ and jammer \\ 
			\end{tabular} \\ \hline
			${s_{_{_{{S_T}}}}}, {s_{_{_{{S_J}}}}}$ &
            \begin{tabular}{l}
			 Signals sent by BS and jammer \\ 
			\end{tabular} \\ \hline
			$\gamma$ & 
            \begin{tabular}{l}
			 SJNR \\ 
			\end{tabular} \\ \hline
			${\sigma ^2}$	 & 
             \begin{tabular}{l}
			 Noise \\ 
			\end{tabular} \\ \hline		
		\end{tabular}
	\end{center}
\end{table}

\subsection{Main Contributions and Outcomes}
Thus far, despite significant advancements in mitigating jamming in RIS enabled systems, the existing literature has not investigated jamming on satellites. This gap is particularly critical because satellite- based jamming poses unique challenges due to its extensive coverage and persistent LoS communication capabilities, disrupting communication across the space, aerial, and ground layers simultaneously. The main contributions of this paper are given as follows:

\begin{enumerate}
     \item This work investigates a challenging and rarely explored scenario: unlike ground-based jammers, jamming from a satellite can cause severe interference due to its high altitude and wide coverage. To the best of our knowledge, no prior work has considered this scenario, making it a novel and important direction to explore. To this study, we explore the static of downlink (DL) transmission scenario where a Low Earth Orbit (LEO) satellite serves as the primary transmitter, a GEO satellite acts as a jammer, and a RIS is mounted on a fixed Unmanned Aerial Vehicle (UAV) to assist in communication. The UAV enables flexible RIS placement, ensuring strong LoS with the LEO satellite. This positioning helps improve signal quality and suppress jamming, advantages that are often limited with ground-mounted RIS due to obstacles and the topology of buildings.
% To this study, we explore the static of downlink (DL) transmission scenario where a Low Earth Orbit (LEO) satellite serves as the primary transmitter, a GEO satellite acts as a jammer, and a RIS is mounted on a fixed Unmanned Aerial Vehicle (UAV) to assist in communication. It should be noted that because of its adaptable placement, the UAV is utilized for RIS deployment. Unlike ground-based RIS, which is limited by building placement and topography, a UAV can be positioned to maintain optimal LoS with the LEO satellite. This ensures the RIS is correctly oriented for maximum signal enhancement and jamming suppression, which may not always be possible with building-mounted RIS.            

     \item We formulate a joint optimization problem for LEO satellite power allocation and RIS passive beamforming, aiming to maximize the signal-to-jamming-plus-noise ratio (SJNR) at the User Equipment (UE). Because the objective function is non-convex, the formulated problem is hard to solve. To make sure to address this issue, an alternating optimization (AO) approach is employed to break the problem into small subproblems, and semidefinite relaxation (SDR) is used to convert the non-convex problem into a convex form.  
    
    \item The results emphasize the impact of the GEO-sourced jamming and highlight the considerable path loss with high-altitude deployed RIS. In addition, the results demonstrate that deploying RIS-equipped UAVs closer to the UE is more effective in mitigating jamming, as proximity reduces path loss and improves signal resilience.
\end{enumerate}

The rest of this article is organized as follows. In Sec. \ref{Sec. System Model} the system model is outlined and introduced the key parameters. The mathematical expressions of optimization are analyzed in Sec. \ref{Sec. Analysis}. Numerical results are presented in Sec. \ref{Sec. Results}. Ultimately,  Sec. \ref{Sec. Conclusion} concludes this article.

\textit{Notations:} In this context, boldface lowercase letters represent vectors, while boldface uppercase letters represent matrices. $diag(.)$ and $tr(.)$ denote a diagonal and trace matrices. The distribution of a complex Gaussian random vector with a mean of $0$ and a variance of $\delta^{2}$ is denoted as $\mathit{CN}(0; \delta^{2})$. The notations of this paper are summarized in Table (\ref{table:Notation}).

% We aim to maximize the signal-to-jamming noise ratio (SJNR) from the LEO satellite to User Equipment (UE) via the joint design of LEO's power allocation and RIS's passive beamforming. The problem we have formulated gives a challenge to solve, primarily because of the non-convex of the objective function. To address this difficulty, we propose the use of a semidefinite relaxation (SDR) technique. Based on simulations, it is shown that when the IRS is deployed with optimized phase beamforming, the SJNR is significantly enhanced rather to when RIS-based systems are used without optimization. Furthermore,  the results illustrate that because of the severe jamming from the GEO satellite and considerable path loss from RIS at high elevations, it is more efficient to deploy the RIS on UAVs near the user.

\section{System Model}\label{Sec. System Model}

As illustrated in Fig.\ref{fig: system_model}, this paper focuses on an anti-jamming communication system. The core elements consist of a LEO satellite functioning as the BS, denoted by ${S_T}$, equipped with a directional antenna. The UE signifies end-user communication devices, while the GEO satellite as the malevolent jammer, denoted by ${S_J}$, endeavors to disrupt ground communication nodes. Unlike ground-based RIS, which is constrained by fixed infrastructure, the RIS in this work is mounted on a UAV to ensure optimal LoS with the LEO satellite, maximizing signal enhancement and jamming suppression. Consequently, a RIS-UAV assisted node, denoted by ${R_U}$, operates aerially to enhance anti-jamming communication and optimize signal transmission. To maintain stable system conditions where all network parameters, including channel states, remain unchanged, the UAV is assumed to be in a fixed position throughout the transmission process.

% \begin{figure}[t] 
%     \centering
%     \includegraphics[width=0.8\linewidth, height=9cm, keepaspectratio=false]{JPG//Cropped_GEO_LEO_UAV_v2.png} %System_Model.png, GEO_LEO_UAV.png
%     \caption{\scriptsize System model consisting of GEO as a jammer ($S_J$), LEO as a transmitter ($S_T$), and RIS mounted on a fixed UAV ($R_U$).}
%     \label{fig: system_model}
% \end{figure}

Each connected link is located in 3D Cartesian coordinates. Therefore, the positions of the transmitter, receiver, jammer, and UAV are considered fixed and expressed by ${{\mathbf{C}}_{{S_T}}} = [{x_{_{{S_T}}}}, {y_{_{{S_T}}}}, {z_{_{{S_T}}}}]$, ${{\mathbf{C}}_{UE}} = [{x_{_{UE}}}, {y_{_{UE}}}, 0]$, ${{\mathbf{C}}_{{{\mathrm{S}}_J}}} = [{x_{_{{S_J}}}}, {y_{_{{S_J}}}}, {z_{_{{S_J}}}}]$, and ${{\mathbf{C}}_{{R_U}}} = [{x_{_{{R_U}}}}, {y_{_{{R_U}}}},{z_{_{{R_U}}}}]$, respectively.

The RIS is equipped with a uniform planar array of $K = {K_r}\times{K_c}$ reflecting elements, accompanied by a smart controller capable of manipulating the reflection phase. The positioning of the RIS is assumed to coincide with the UAV.

Moreover, the diagonal phase shift matrix of the RIS is denoted as ${\mathbf{\Theta}} = \text{diag}({e^{j{\theta _{1}}}}, \ldots,{e^{j{\theta _{K}}}})\in {\mathbb{C}^{K \times K}}$, where  ${\theta _{k}}  \in [0,2\pi )$ is the phase shift of the $k$-th reflecting element with assuming  ${k} \in \{ 1, \ldots, K\} $. Furthermore, considering the continuous controllability of ${\theta _k}$, we disregard signals that undergo multiple reflections from the RIS. 

In this study, we assume all channels within the system are LoS channels. Therefore, the channel gain of the $S_T$ and UE link is described as follows

\begin{equation}\label{eq: Channel LEO-UE}
    {h_{_{{S_T}-UE}}} = \sqrt {\rho d_{_{{S_T}-UE}}^{ - \alpha_d}} {g_{_{{S_T}-UE}}}.
\end{equation}
where, ${\rho d_{_{{S_T}-UE}}^{ -\alpha_d}}$ and ${g_{_{{S_T}-UE}}} = {e^{ - j\frac{{2\pi }}{\lambda }{d_{_{{S_T}-UE}}}}}$ represent the pathloss and phase response, respectively. Furthermore, $\rho$ is the path loss at the reference distance of one meter, and $\alpha_d$ is the path loss exponent between the $S_T$ and UE.
$\lambda$ is the wavelength, and ${{d_{_{{S_T}-UE}}}}$ illustrates the distance between $S_T$ and UE. For the channel from the jammer $S_J$ to UE, i.e. ${h_{_{{S_J}-UE}}}$, the same model is used, as given as follows

\begin{equation}\label{eq: Channel LEO-UE}
    {h_{_{{S_J}-UE}}} = \sqrt {\rho d_{_{{S_J}-UE}}^{ - \alpha_d}} {g_{_{{S_J}-UE}}}.
\end{equation}
where, the ${\rho d_{_{{S_J}-UE}}^{ -\alpha_d}}$ and ${g_{_{{S_J}-UE}}} = {e^{ - j\frac{{2\pi }}{\lambda }{d_{_{{S_J}-UE}}}}}$ represent the pathloss and phase response between $S_J$ and UE, respectively.

In addition, the  $\text S_T$-$\text R_U$ channel, RIS's reflection with phase shifts, and $\text R_U$-UE channel are three parts that are combined to model the $\text S_T$-$\text R_U$-UE channel. 

In particular, the $\text S_T$-$\text R_U$ channel indicates by \cite{RIS_beamformin_DiRenzo}

\begin{equation} \label{eq: Channel LEO-RIS}
  {\bold{h_{_{{S_T}-{R_U}}}}} = \sqrt {\rho d_{_{{S_T}-{R_U}}}^{ - \alpha_r}} {\bold{g_{_{{S_T}-{R_U}}}} }.
\end{equation}
where, ${\rho d_{_{{S_T}-{R_U}}}^{ - \alpha_r}}$ denotes the pathloss between $\text S_T$ and RIS. Moreover, ${d_{_{{S_T}-{R_U}}}}$ represents the distance from $\text S_T$ to the RIS. Furthermore, the phase response of the $\text S_T$-RIS channel is provided by
\begin{equation}\label{eq: Channel LEO-RIS_LoS} 
\begin{split}
{\bold {g_{_{{S_T}-{R_U}}}}} &= {[1, \ldots ,{e^{ - j\frac{{2\pi }}{\lambda }d({K_r} - 1){a_x}}}]^T} \\
& \otimes {[1, \ldots ,{e^{ - j\frac{{2\pi }}{\lambda }d({K_c} - 1){a_z}}}]^T},
\end{split}
\end{equation}
where 
\begin{equation*}
   {a_x} = \sin {\phi _{_{{S_T}-{R_U}}}}\cos {\varphi _{_{{S_T}-{R_U}}}}.
\end{equation*}
\begin{equation*}
   {a_z} = \sin {\phi _{_{{S_T}-{R_U}}}}\sin {\varphi _{_{{S_T}-{R_U}}}}.
\end{equation*}

Also, $d$ is the separation of the RIS element. Additionally, ${\phi _{_{{S_T}-{R_U}}}}$ and ${\varphi_{_{{S_T}-{R_U}}}}$ denote the angle of arrival (AoA) in both vertical and horizontal directions, respectively, with:
\begin{equation*}
\begin{split}
  \sin {\phi_{_{{S_T}-{R_U}}}} &= \frac{{\left|{z_{_{R_U}}} - {z_{_{{S_T}}}}\right|}}{{{d_{_{{S_T}-{R_U}}}}}},\\
   \sin {\varphi _{_{{S_T}-{R_U}}}} &= \frac{{\left| {x_{_{R_U}}} - {x_{_{{S_T}}}}\right|}}{{\sqrt {{{\left( {{x_{_{R_U}}} - {x_{_{{S_T}}}}} \right)}^2} + {{\left( {{y_{_{R_U}}} - {y_{_{{S_T}}}}} \right)}^2}} }},\\
   \cos {\varphi _{_{{S_T}-{R_U}}}} &= \frac{{\left| {y_{_{R_U}}} - {y_{_{{S_T}}}}\right|}}{{\sqrt {{{\left( {{x_{_{R_U}}} - {x_{_{{S_T}}}}} \right)}^2} + {{\left( {{y_{_{R_U}}} - {y_{_{{S_T}}}}} \right)}^2}} }}.
\end{split}
\end{equation*}

A similar method is employed to represent the RIS-UE channel, i.e. $\bold{h_{_{{R_U}-UE}}}$. Therefore, the whole channel of ${\text S_T}$-${R_U}$-UE is expressed by 
\begin{equation}
    {h_{_{{S_{T}}-{R_U}-UE}}} = \bold {h_{_{{R_U}-UE}^{}}^H}\bold {\Theta }\bold {{h_{_{{S_T}-{R_U}}}}}.
\end{equation}

%A similar approach is valid for obtaining the channel of ${\text S_J}$-${R_U}$-UE. 
Similarly, the indirect channel between the jammer $S_J$ and the user, UE, is  given as
\begin{equation}
    {h_{_{{S_{J}}-{R_U}-UE}}} = \bold {h_{_{{R_U}-UE}^{}}^H}\bold {\Theta }\bold {{h_{_{{S_J}-{R_U}}}}},
\end{equation}
where $\bold {h_{_{{R_U}-UE}^{}}^H}$ and $\bold {{h_{_{{S_J}-{R_U}}}}}$ are the channel gains of the links from $S_J$ to $R_U$ and from $R_U$ to UE, respectively. 
Therefore, the signal received by the UE is given as

\begin{equation}\label{eq:received signal} 
\begin{split}
y =& \sqrt {{P_{{{{S_T}}}}}} \left( {{h_{_{{S_{T}}-UE}}} + {{h_{_{{S_{T}}-{R_U}-UE}}}}} \right){{s_{_{_{{S_T}}}}}} + \\
&\sqrt {{P_{{{{S_J}}}}}} \left( {{h_{_{{S_{J}}-UE}}} + {{h_{_{{S_{J}}-{R_U}-UE}}}}} \right){{s_{_{_{{S_J}}}}}} + {\mathrm{w}},
\end{split}
\end{equation}
where, ${P_{{{{S_T}}}}}$ and ${P_{{{{S_J}}}}}$ illustrate the transmit power of the LEO satellite as a BS and GEO satellite as a jammer. ${s_{_{_{{S_T}}}}}$ and ${s_{_{_{{S_J}}}}}$ denote the signals sent by BS and jammer. In addition, ${\mathrm{w}}$ points Additive White Gaussian Noise (AWGN) with zero mean and ${\sigma ^2}$ variance.

Accordingly, the SJNR is calculated as follows 
\begin{equation} 
\gamma = \frac{{{P_{{{{S_T}}}}}{{\left| {{h_{_{{S_{T-}}UE}}} + {{h_{_{{S_{T}}-{R_U}-UE}}}}}  \right|}^2}}}{{{P_{{{{S_J}}}}}{{\left| {{h_{_{{S_{J-}}UE}}} + {{h_{_{{S_{J}}-{R_U}-UE}}}}} \right|}^2} + {\sigma ^2}}}.
\end{equation}

\section{Analysis and Problem Formulation } \label{Sec. Analysis}
According to the provided system model, our goal is to maximize the SJNR by jointly optimizing the phase shift matrix of RIS and the transmit power of the LEO satellite under power constraints. Moreover, the RIS is assumed to have access to channel state information (CSI). This idealized assumption enables a clear understanding of the system's performance limits and serves as a useful baseline for more practical scenarios. Mathematically, the problem can be formulated as follows
\vspace{-0.1cm}
\begin{subequations} 
\begin{align} \label{eq: problem formulation}
% \begin{array}{l}
\text{(P0):} &\mathop {\max }\limits_{\mathbf{\Theta}, {P_{{{{S_T}}}}}} {\gamma} \\
\text{s.t.}\quad &{\mathrm{ }}{\theta_{k}}{\mathrm{ }} \in {\mathrm{[0}}{\mathrm{,2}}\pi {\mathrm{)}}, \forall k \label{eq: shift phase constraint}\\
& 0 \le {P_{{{{S_T}}}}} \le {P_{{max_{{S_T}}}}} \label{eq: max power constraint}
\end{align}
\end{subequations}
where, ${P_{{max_{{S_T}}}}}$ is the maximum transmit power of the LEO satellite. The restriction on the phase shift for each reflective element is presented in Eq. (\ref{eq: shift phase constraint}). The defined problem under these constraints is not convex and at this stage, there is no exact approach to solve the problem. In the following, we can apply the AO method. This approach alternates between two sub-problems:

\begin{enumerate}
    \item Optimize the transmit power ${P_{{{{S_T}}}}}$ while keeping the RIS phase matrix fixed.
    \item Optimize the RIS phase matrix while keeping ${P_{{{{S_T}}}}}$ fixed.
\end{enumerate}

% These steps are iterated until convergence is achieved.

% \section{Analysis}\label{Sec. Analysis}

\subsection{Optimizing ${P_{{{{S_T}}}}}$ for a given $\Theta $} \label{Sub sec. Optimizing P}

For the given RIS phase shift matrix $\bold \Theta $, the optimization problem in ({\ref{eq: problem formulation}}) can be described  by

\begin{equation}\label{eq: optimization power, fixed theta}
\begin{split}
      &\mathop {\max }\limits_{{P_{S_T}}}\quad \frac{{P_{S_T}}\Gamma}{{P_{S_J}}\Delta + {\sigma ^2}}\\
&\text{s.t.}\quad 0 \le {P_{{{{S_T}}}}} \le {P_{{max_{{S_T}}}}}
\end{split},
\end{equation}
where, $\Gamma = {\left| {{h_{_{{S_T}-UE}}} + {\mathbf{{h_{_{{R_U}-UE}^{}}^H} {\Theta} {{h_{_{{S_T}-{R_U}}}}}}}} \right|}^2$, $\Delta={\left| {{h_{_{{S_J}-UE}}} + {\mathbf{{h_{_{{R_U}-UE}^{}}^H} {\Theta} {{h_{_{{S_J}-{R_U}}}}}}}} \right|}^2$ represent the total  channel gains between $S_T$ and UE and between $S_J$ and $UE$, respectivel 

Taking the derivative of the objective function with respect to ${P_{S_T}}$ and setting it to zero leads to ${P^*_{S_T}}=P_{maxS_T}$. Since $\Gamma$ is constant with respect to ${P_{S_T}}$, so the optimal value is the maximum allowable power.
\subsection{Optimizing $\Theta $ for given ${P_{S_T}}$} \label{Sub sec. Optimizing Theta}
 With ${P_{{{{S_T}}}}}$ fixed, the problem in ({\ref{eq: problem formulation}}) becomes:

\begin{equation}
\begin{split}\label{eq:optimization_theta_fixed_power}
&\mathop {\max }\limits_{\mathbf{\Theta}} \quad \frac{{P_{{S_T}} \left| h_{{S_T}-UE} + \mathbf{h_{{R_U}-UE}^H \Theta h_{{S_T}-{R_U}}} \right|^2}}{{P_{{S_J}} \left| h_{{S_J}-UE} + \mathbf{h_{{R_U}-UE}^H \Theta h_{{S_J}-{R_U}}} \right|^2 + \sigma^2}} \\
&\text{s.t.} \qquad \theta_k \in [0, 2\pi), \forall k 
\end{split}
\end{equation}

For given transmit power  ${P_{{{{S_T}}}}}$, by considering the following equation
\begin{equation}
\bold g_\chi ^H {U_\chi }\upsilon = {h_{_{{\chi }-UE}}} + {h_{_{{\chi }-R_U-UE}}} 
\end{equation}
where,
\begin{align*}
   {\bold U_\chi } &= diag\left( {\left[ {\bold {h_{_{R_U-UE}}}\quad{h_{_{{\chi }-UE}}}} \right]} \right) \\ 
  {\bold g_\chi }  &= {\left[ {\bold h_{_{{S_\chi }-R_U}}^H, 1} \right]^H}, \quad {\chi } \in \left\{ {{S_T},{S_J}} \right\} \\
   {\boldsymbol{\upsilon}} &= [{e^{j{\theta _1}}}, \ldots ,{e^{j{\theta _K}}},1]
\end{align*}

Therefore, Eq. (\ref{eq: problem formulation}) can be changed into

\begin{gather}\label{eq: SDR}
% \begin{array}{l}
\mathop {\max }\limits_\upsilon  {\left( { \frac{{{P_{{{{S_T}}}}}{{\left| {{\bf{g}}_{{S_T}}^H\bf{U_{{S_T}}}\boldsymbol\upsilon} \right|}^2}}}{{{P_{{{S_J}}}}{{\left| {{\bf{g}}_{{S_J}}^H\bf{U_{{S_J}}}\boldsymbol\upsilon} \right|}^2} + {\sigma ^2}}}} \right)} \\
\text{s.t.} \quad \theta_k \in [0, 2\pi), \forall k s\nonumber
\end{gather}

Furthermore, we transform the problem by expressing  
\begin{equation}
  {\left| {\bold g_\chi ^H{\bold U_\chi } \boldsymbol \upsilon } \right|^2} = {\left|{\mathop{\mathrm tr}\nolimits} \left( {{\bold D_\chi }{\mathbf{V}}} \right) \right|}
\end{equation}
where ${\bold D_\chi } = \bold U_\chi ^H{{\bf{g}}_\chi }{\bf{g}}_\chi ^H{\bold U_\chi }$, ${\bf{V}} = \boldsymbol \upsilon {\boldsymbol \upsilon ^H}$ (a positive semidefinite matrix) and tr denotes the trace of a matrix. The problem becomes:

\begin{gather}\label{eq: SDP}
% \begin{array}{l}
\mathop {\max }\limits_V{\left( { \frac{{{P_{{{{S_T}}}}}{{\left|{\mathop{\mathrm tr}\nolimits} \left( {{\mathbf{D_{S_T}}}{\mathbf{V}}} \right) \right|}}}}{{{P_{{{S_J}}}}{{\left|{\mathop{\mathrm tr}\nolimits} \left( {{\mathbf{D_{S_J}}}{\mathbf{V}}} \right) \right|}} + {\sigma ^2}}}} \right)} \\
\text{s.t.} \quad  \mathbf{V}\succeq 0,\nonumber\\
rank (\mathbf{V}) = 1 \nonumber
\end{gather}

The rank constraint $rank(\mathbf{V})=1$ is relaxed to make the problem convex, enabling us to use standard methods for solving SDP. After solving for $\textbf{V}$, we can use eigenvalue decomposition or Gaussian randomization to extract $\boldsymbol\upsilon$. Then reconstruct $\mathbf{\Theta}=diag (\boldsymbol\upsilon)$. \\
In the proposed approach, the objective function remains non-decreasing at each step, ensuring iterative improvement. The stopping criterion is set as either a fixed number of iterations or when the relative change in the objective function falls below a predefined threshold denoted by $\varepsilon$. Since each sub-problem is solved optimally within its domain, the overall optimization process is bounded and converges to a stable solution. In cases where the obtained $\mathbf{V}$ does not satisfy the rank-one constraint, Gaussian randomization is applied to obtain a feasible approximation while preserving performance.

% These steps repeat until convergence or a predefined maximum number of iterations. 
\subsection{Complexity}
 The complexity analysis focuses on the two sub-problems. Since the optimization of transmit power is a scalar problem, its computational complexity is $O(1)$, which is negligible compared to the RIS phase optimization. The RIS phase optimization problem is transformed into a SDP problem via SDR. As a result, the complexity is generally expressed by $O(N^2_d \sqrt{N_c})$, where $N_d$ is the number of decision variables and $N_c$ denotes the number of constraints. By substituting the specific problem parameters, we obtain $O(K^{3.5})$. Consequently, the total computational complexity of this approach is 
$O(IK^{3.5})$, where $I$ is the number of iterations required for solving Eq. (\ref{eq: problem formulation}).  Furthermore, the proposed method also shows stable behavior in terms of convergence. The SJNR value increases or stays constant in each iteration, and convergence is typically reached within a small number of steps.

\section{Numerical Results}\label{Sec. Results}
This section provides numerical results to demonstrate the performance of the proposed method. The parameters are defined as follows: $\mathbf{C}_{\mathrm{S_T}}=[0,0,500]$ Km, $\mathbf{C}_{\mathrm{S_J}}=[0,0,35786]$ Km, ${\mathbf{C}}_{R_U}=[0,0,50]$ m, ${\mathbf{C}}_{UE}=[0,0,0] $ m, $P_{S_T}=20$ dBW, $P_{S_J}=30$ dBW, $BW=1$ MHz, $\text{Noise spectral density}=-174$ dBm/Hz, $\text{Noise Figure}=1$ dB, $\rho = -55$ dB and the stopping threshold for relative improvement in SJNR is set to $\varepsilon =10^{-3}$. 
%These values are chosen based on commonly used settings in similar optimization problems to balance accuracy and computational efficiency.

Fig. \ref{Fig: fix_RIS_50m, various LEO distance} shows the significant role of optimized RIS in improving SJNR compared to non-optimized configurations, with the enhancements becoming more evident as the number of RIS elements increases.  As the LEO satellite distance increases, the SJNR decreases in both cases. This is expected due to the increased path loss resulting from the larger distance. However, by increasing the number of RIS elements, the benefits of optimization become more visible, resulting in notable SJNR gains and improved resilience to path loss over longer distances. Furthermore, the optimized RIS outperforms the non-optimized configurations across all RIS sizes. Even with fewer elements, an optimized RIS can achieve higher SJNR performance by aligning phase shifts effectively, while non-optimized RIS configurations fail to adequately suppress jamming signals and apply constructive interference at the receiver.  

% Overall, the non-optimized RIS provides limited gains due to the random phase shifts, which do not suppress the jamming signals to maximize constructive interference at the receiver.

\begin{figure}[t]
    \centering
    \includegraphics[width=0.475\textwidth, height = 6.5cm]{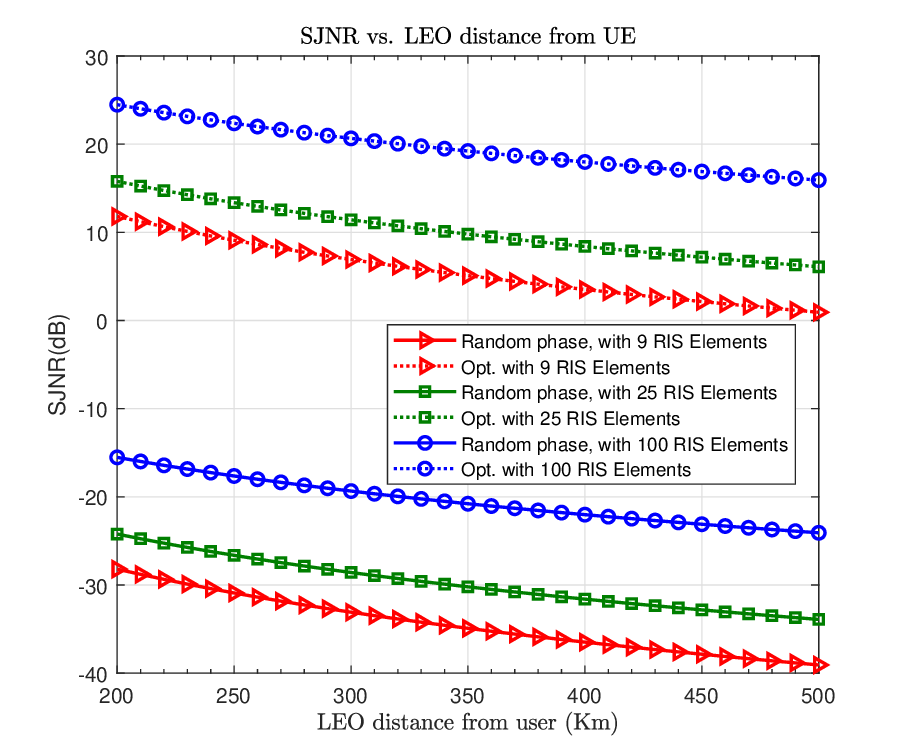}
    \caption{SJNR as a function of LEO distances from UE with $9$, $25$, and $100$ RIS elements where RIS is located at $50 $ m above the ground.}
    \label{Fig: fix_RIS_50m, various LEO distance}
\end{figure}
% \vspace{-3mm}
\begin{figure}[t]
    \centering
    \includegraphics[width=0.475\textwidth, height = 6.5cm]{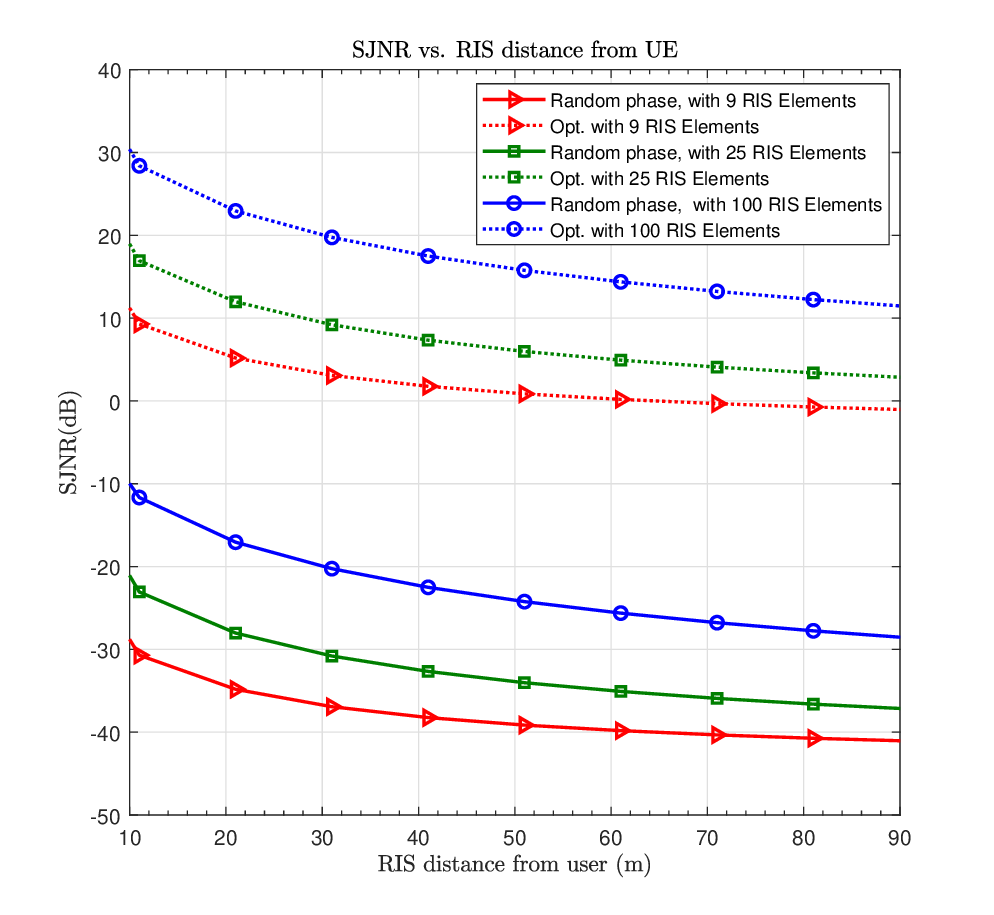}
    \caption{SJNR as a function of RIS distances from UE with $9$, $25$, and $100$ RIS elements where LEO is located at $500$ Km above the ground.}
    \label{Fig:fix_LEO_500km_various_RIS_distances}
\end{figure}
\begin{figure}[t]
    \centering
    \includegraphics[width=0.475\textwidth, height = 6.5cm]{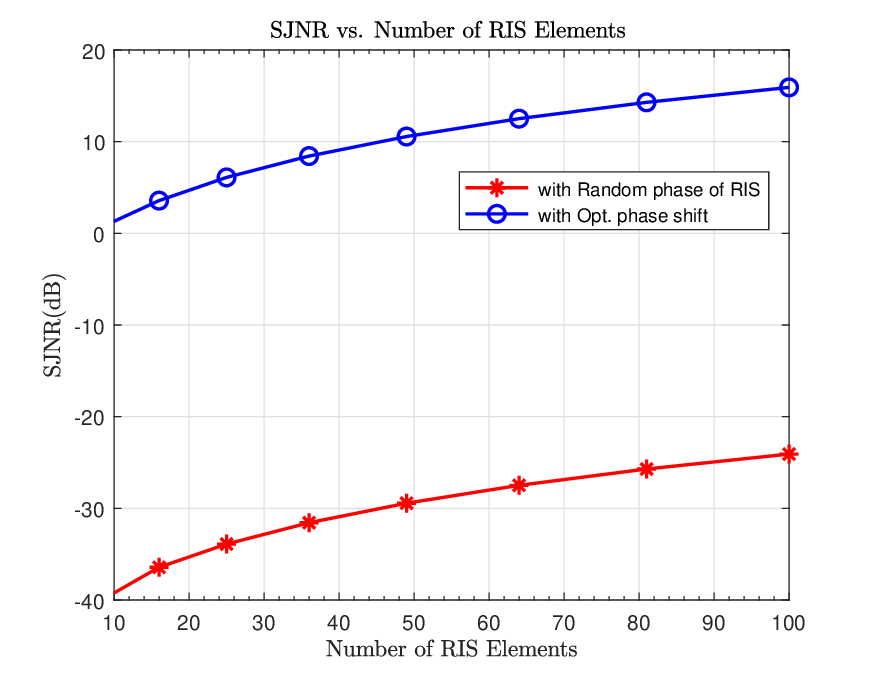}
    \caption{SJNR as a function of number of RIS elements, with $d_{{S_T}-UE}=500 km$, $d_{{R_U}-UE}=50 m$.}
    \label{Fig: Various_RIS_number}
\end{figure}

% \begin{figure} [h]
%      \centering
%      \begin{subfigure}
%          \centering
%          \includegraphics[width=0.4\textwidth]{JPG/9_RIS.eps}
%          % \caption{}
%          \label{Fig:9_RIS}
%      \end{subfigure}
%      % \hfill
%      \begin{subfigure}[a]
%          \centering
%          \includegraphics[width=0.4\textwidth]{JPG/25_RIS.eps}
%          % \caption{}
%          \label{Fig: 25_RIS}
%      \end{subfigure}
%      % \hfill
%       \begin{subfigure}[b]
%          \centering
%          \includegraphics[width=0.4\textwidth]{JPG/100_RIS.eps}
%          % \caption{}
%          \label{Fig:100_RIS}
%      \end{subfigure}[c]
%         \caption{SJNR vs. LEO distances from UE with $9$, $25$, and $100$ RIS elements where RIS is located at $50 $ m above the ground.} 
%          \label{Fig: fix_RIS_50m, various LEO distance}
% \end{figure}

Fig. \ref{Fig:fix_LEO_500km_various_RIS_distances}, we study the impact of RIS distance from UE. For all states, the SJNR decreases as the distance between the RIS and the UE increases. This behavior is expected because, as the RIS moves further from the UE, the reflected signal experiences greater path loss and reduced strength, leading to a drop in SJNR. Similar to previous results, optimized RIS enhances the SJNR performance compared to the non-optimized case. Furthermore, increasing the number of RIS elements improves the SJNR, especially at close distances to the user, due to better reflection and beamforming capabilities. However, as the RIS distance continues to increase, the SJNR diminishes significantly, and it seems that at distances greater than 90 meters, the graphs with different RIS dimensions will overlap. This convergence suggests that the additional elements in larger RIS configurations may no longer provide significant benefits due to excessive path loss and reduced signal power at longer distances. Moreover, the potential overlap or crossing of curves at extreme distances highlights the limits of RIS performance without proper placement and optimization, emphasizing the criticality of proximity to the UE for maintaining effective anti-jamming capabilities.

% \begin{figure}[h]
%     \centering
%     % First subfigure
%     \begin{subfigure}
%         \centering
%         \includegraphics[width=0.4\textwidth]{JPG/9_RIS_various_distance.eps}
%         % \caption{}
%         \label{Fig:9_RIS}
%     \end{subfigure}[a]
%     \vspace{0.5em} % Vertical space between subfigures
%     % Second subfigure
%     \begin{subfigure}
%         \centering
%         \includegraphics[width=0.4\textwidth]{JPG/25_RIS_various_distance.eps}
%         % \caption{}
%         \label{Fig:25_RIS}
%     \end{subfigure}
%     \vspace{0.5em} % Vertical space between subfigures
%     % Third subfigure
%     \begin{subfigure}[b]
%         \centering
%         \includegraphics[width=0.4\textwidth]{JPG/100_RIS_various_distance.eps}
%         % \caption{}
%         \label{Fig:100_RIS}
%     \end{subfigure}[c]
%     \caption{SJNR vs. RIS distances from UE with $9$, $25$, and $100$ RIS elements where LEO is located at $500$ Km above the ground.}
%     \label{Fig:fix_LEO_500km_various_RIS_distances}
% \end{figure}
\vspace{0.1cm}
In Fig. \ref{Fig: Various_RIS_number}, optimization improves the beamforming capability of the RIS, which efficiently focuses the reflected signal toward the desired direction while mitigating the jamming signal. The gap between the two curves widens as the number of RIS elements increases. This demonstrates that the benefit of optimization becomes more pronounced when the RIS has more elements, as it allows for better beamforming and signal enhancement.

While the results mainly focus on the effect of RIS altitude, the horizontal location of the RIS also plays an important role. In a practical setup, the total distance between the RIS and the user, as well as the signal reflection angles, change when the RIS is moved horizontally. These changes affect the overall signal strength, especially when the RIS is far from the user or poorly aligned. Our LoS-based model suggests that the best performance is achieved when the RIS is placed close to the user and properly oriented. This highlights one of the main advantages of using a UAV to carry the RIS: unlike fixed structures such as buildings, a UAV can be positioned more flexibly to achieve both vertical and horizontal alignment. Although this aspect is not directly shown in the figures, it is an important design factor and will be explored in future work.

% \FloatBarrier

%%%%%%%%%%%%%%%
\section{Conclusion} \label{Sec. Conclusion}
In this paper, we studied a challenging and underexplored anti-jamming scenario in which a GEO satellite acts as a jammer targeting a SAGIN downlink. Unlike terrestrial jamming, GEO-originated attacks represent a worst-case threat due to persistent LoS and wide-area coverage. To address this, we employed a UAV-mounted RIS to enhance signal quality and suppress interference. By jointly optimizing RIS beamforming and LEO satellite transmit power, we demonstrated significant improvements in SJNR. Our results also highlight the importance of RIS placement, with closer proximity to ground users yielding better performance. While ideal conditions such as perfect CSI and fixed UAV positioning were assumed, future work will address UAV mobility, time-varying channels, and energy constraints to reflect more practical environments. This study provides a foundation for designing resilient communication strategies against satellite-based jamming in SAGIN.
%  In this paper, we addressed the critical challenge of mitigating jamming in SAGIN using RIS mounted on a UAV.
% By optimizing RIS beamforming and the LEO satellite's transmit power, significant improvements in SJNR were achieved. Numerical results validate that optimized RIS configurations outperform non-optimized setups, especially when the RIS is deployed closer to the ground user, reducing path loss and enhancing resilience against high-powered jamming from GEO satellites. This research highlights the potential of RIS-based anti-jamming techniques as a transformative approach to improving SAGIN's communication quality and reliability under adversarial conditions. This study, instead of focusing on a new algorithm, aims to evaluate the effectiveness of UAV-mounted RIS in a satellite jamming scenario, which has not been well explored in the literature. These findings offer a helpful starting point, even though we simplified the research by assuming ideal conditions. More realistic configurations, such as mobile UAVs, imperfect CSI, and energy constraints, can be investigated in future research. Overall, our proposed method is simple and flexible and has the potential to improve the resilience of satellite communication networks.
%\section{Appendix}\label{Appendix}
% \IEEEQED
\vspace*{-5mm}
%\begin{thebibliography}{9}
%	

\bibliographystyle{IEEEtran}
\bibliography{Ref}
\end{document}